\documentclass[twocolumn,showpacs,preprintnumbers,amsmath,amssymb,nofootinbib]{revtex4}
\usepackage{amsmath}
\usepackage{graphicx}
\usepackage{calc}
\usepackage{bm}
\usepackage{float}

\begin{document}
\newcommand{\beq}{\begin{equation}}
\newcommand{\eeq}{\end{equation}}

\title{Spin and valley effects on the quantum phase transition in two dimensions}
\author{A.~A. Shashkin$^a$ and S. V. Kravchenko$^b$\vspace{2mm}}
\affiliation{$^a$Institute of Solid State Physics, Chernogolovka, Moscow District, 142432, Russia\\
$^b$Physics Department, Northeastern University, Boston, Massachusetts 02115, USA}

\textit{Contribution to JETP special issue in honor of E.~I.~Rashba's 95th birthday}

\begin{abstract}
Using several independent methods, we find that the metal-insulator transition occurs in the strongly-interacting two-valley two-dimensional electron system in ultra-high mobility SiGe/Si/SiGe quantum wells in zero magnetic field. The transition survives in this system in parallel magnetic fields strong enough to completely polarize the electrons' spins, thus making the electron system ``spinless''. In both cases, the resistivity on the metallic side near the transition increases with decreasing temperature, reaches a maximum at a temperature $T_{\text{max}}$, and then decreases. The decrease reaches more than an order of magnitude in zero magnetic field. The value of $T_{\text{max}}$ in zero magnetic field is found to be close to the renormalized Fermi temperature. However, rather than increasing along with the Fermi temperature, the value $T_{\text{max}}$ decreases appreciably for spinless electrons in spin-polarizing magnetic fields. The observed behavior of $T_{\text{max}}$ cannot be described by existing theories. The results indicate the spin-related origin of the effect. At the same time, the low-temperature resistivity drop in both spin-unpolarized and spinless electron systems is described quantitatively by the dynamical mean-field theory.
\end{abstract}
\maketitle

\date{\today}

\section{Introduction}

Spin and valley degrees of freedom in two-dimensional (2D) electron systems have recently attracted much attention due to rapidly developing fields of spintronics and valleytronics (see, \textit{e.g.}, Refs.~\cite{behnia2012polarized,zhu2012field,schaibley2016valleytronics,zhu2017emptying}).  The existence of the zero-magnetic-field metallic state and the metal-insulator transition (MIT) in strongly interacting 2D electron systems is intimately related to the existence of these degrees of freedom \cite{lee1985disordered,punnoose2001dilute,punnoose2005metal,fleury2008many}.  The MIT in two dimensions was theoretically envisioned based on the renormalization group analysis (see Ref.~\cite{lee1985disordered} for a review).  It was first experimentally observed in a strongly-interacting 2D electron system in silicon metal-oxide-semiconductor field-effect transistors (MOSFETs) \cite{zavaritskaya1987metal,kravchenko1994possible,kravchenko1995scaling, popovic1997metal} and subsequently reported in a wide variety of 2D electron and hole systems: $p$-type SiGe heterostructures, $p$- and $n$-type GaAs/AlGaAs heterostructures, AlAs heterostructures, ZnO-related heterostructures, \textit{etc}.\ (for recent reviews, see Refs.~\cite{shashkin2019recent,shashkin2021metal}). Now it is widely accepted that the driving force behind the MIT is the strong correlations between carriers. Here we study the metal-insulator transition and non-monotonic temperature-dependent resistivity on the metallic side near the MIT in the strongly-interacting two-valley 2D electron system in ultra-high mobility SiGe/Si/SiGe quantum wells in zero and spin-polarizing magnetic fields.

Measurements reported here were performed on ultra-high mobility SiGe/Si/SiGe quantum wells similar to those described in Refs.~\cite{melnikov2015ultra,melnikov2017unusual}. The peak electron mobility, $\mu$, in these samples reaches 240~m$^2$/Vs. It is important to note that judging by the appreciably higher quantum electron mobility ($\sim10$~m$^2$/Vs) in the SiGe/Si/SiGe quantum wells compared to that in Si MOSFETs, the residual disorder related to both short- and long-range random potential is drastically smaller in the samples used here. The approximately 15~nm wide silicon (001) quantum well is sandwiched between Si$_{0.8}$Ge$_{0.2}$ potential barriers. The samples were patterned in Hall-bar shapes with the distance between the potential probes of 150~$\mu$m and width of 50~$\mu$m using standard photo-lithography. Measurements were carried out in an Oxford TLM-400 dilution refrigerator. Data on the metallic side of the transition were taken by a standard four-terminal lock-in technique in a frequency range 1--10~Hz in the linear response regime. On the insulating side of the transition, the resistance was measured with {\it dc} using a high input impedance electrometer. Since in this regime, the current-voltage ($I$-$V$) curves are strongly nonlinear, the resistivity was determined from ${\rm d}V/{\rm d}I$ in the linear interval of $I$-$V$ curves, as $I\rightarrow0$.

\section{Quantum phase transition in zero magnetic field}

An important metric defining the MIT is the magnitude of the resistance drop in the metallic regime. Until recently, the strongest drop of the resistance with decreasing temperature (up to a factor of 7) was reported in Si MOSFETs \cite{kravchenko1994possible,kravchenko1995scaling}. In contrast, in spite of much lower level of disorder in $p$- and $n$-GaAs-based structures, the low-temperature drop of the resistance in those systems has never exceeded a factor of about three \cite{hanein1998the}. This discrepancy has been attributed primarily to the fact that electrons in Si MOSFETs have two almost degenerate valleys, which further enhances the correlation effects \cite{punnoose2001dilute,punnoose2005metal}. The importance of these strong interactions in 2D systems has been confirmed recently in the observation of the formation of a quantum electron solid in Si MOSFETs \cite{brussarski2018transport}. It has been found that the effective electron mass in Si MOSFET 2D electron systems strongly increases as the electron density is decreased, with a tendency to diverge at a density that lies close to, but is consistently below, the critical density for the MIT \cite{shashkin2002sharp,mokashi2012critical}.

\begin{figure}
\scalebox{.95}{\includegraphics[width=\columnwidth]{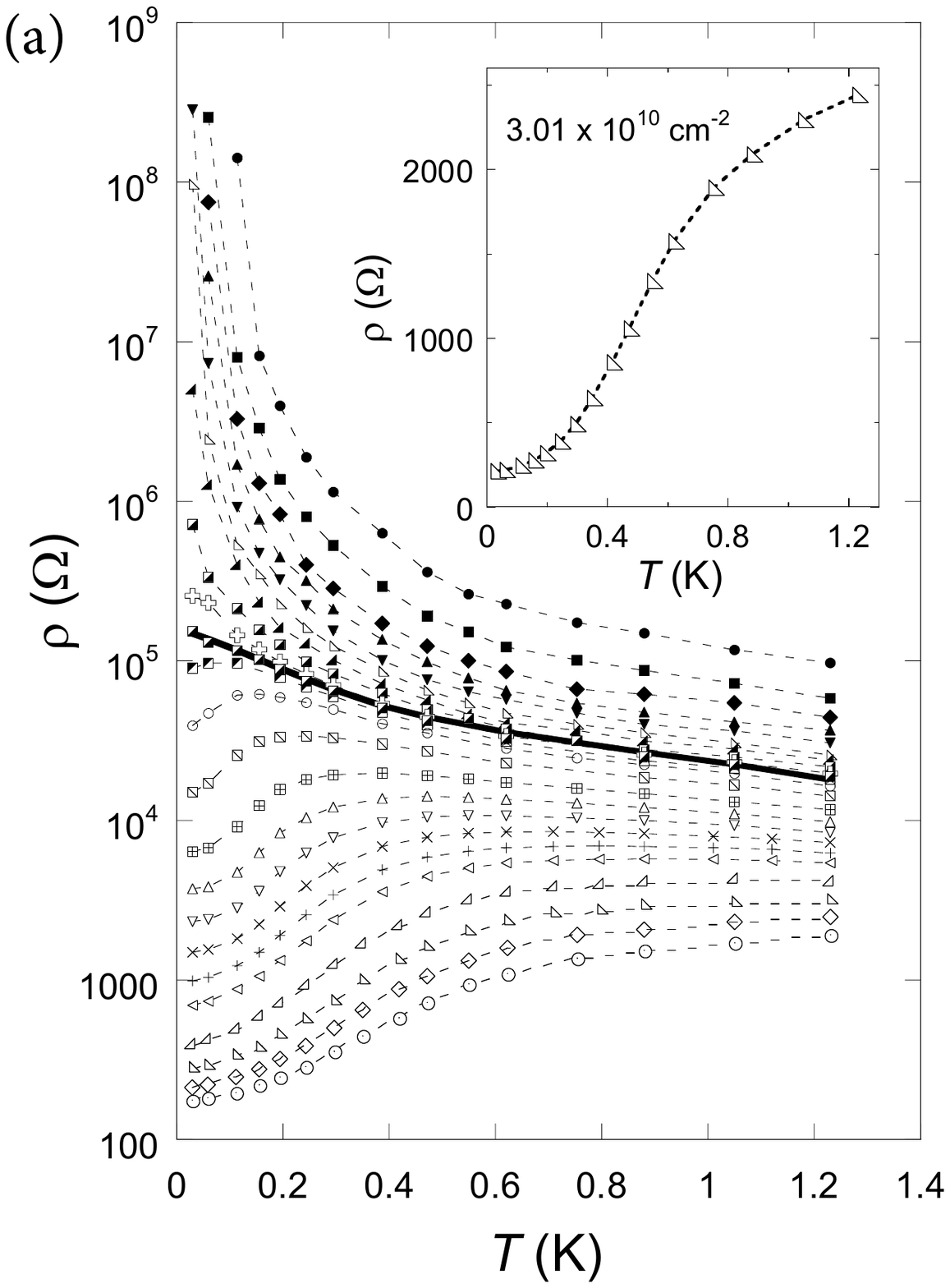}}
\scalebox{.95}{\includegraphics[width=\columnwidth]{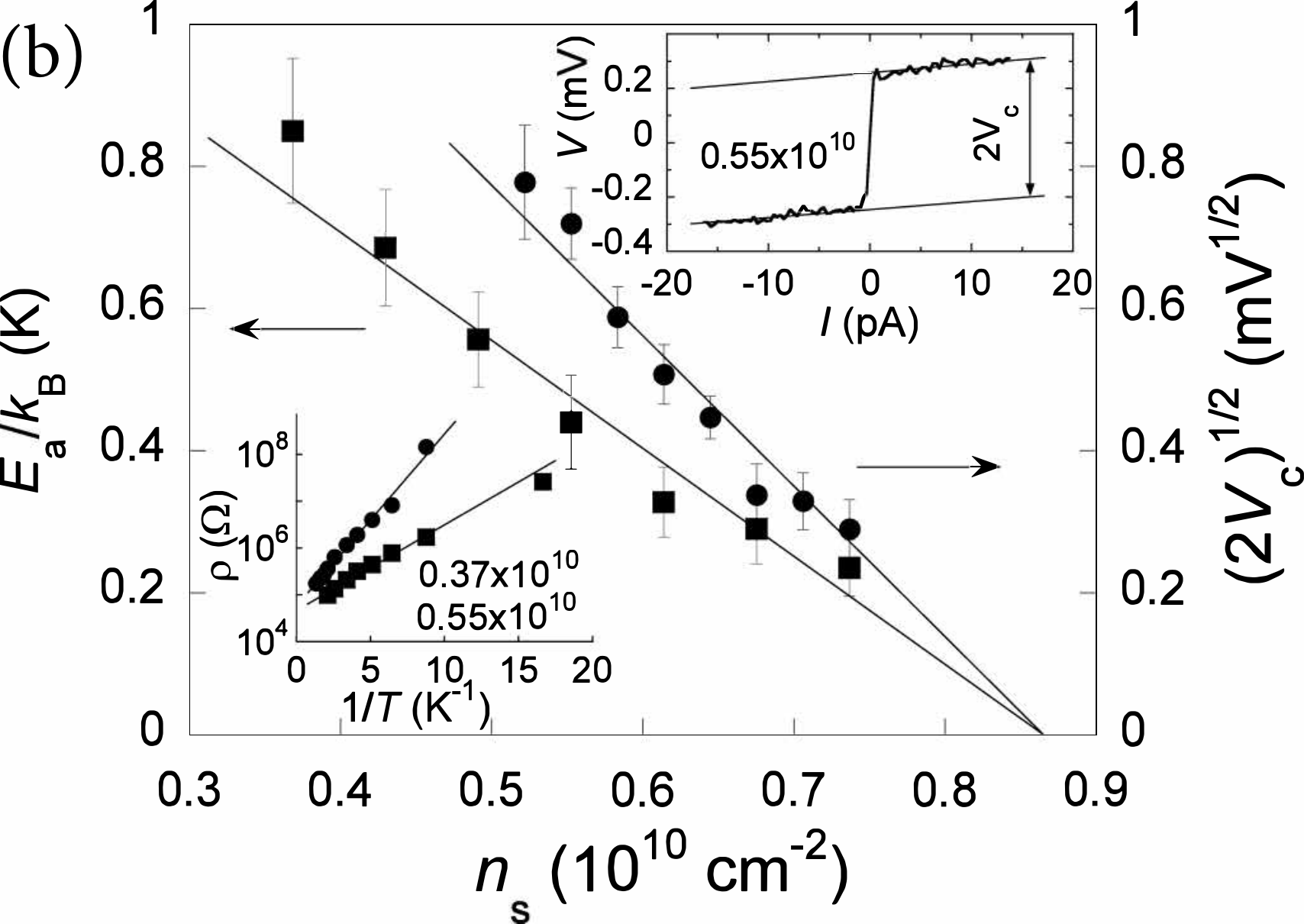}}
\caption{(a) Temperature dependences of the resistivity in a SiGe/Si/SiGe quantum well in zero magnetic field. The electron densities in units of $10^{10}$~cm$^{-2}$ (top to bottom) are: 0.37, 0.43, 0.49, 0.55, 0.61, 0.68, 0.74, 0.80, 0.85, 0.88, 0.92, 0.98, 1.17, 1.35, 1.54, 1.72, 1.90, 2.09, 2.27, 2.64, 3.01, 3.38, and 3.75. The solid line corresponds to the separatrix. The inset shows a close-up view of $\rho(T)$ at $n_{\text s}=3.01\times10^{10}$~cm$^{-2}$ displaying a drop of the resistivity by a factor of 12. (b) Activation energy and the square root of the threshold voltage as a function of the electron density in zero magnetic field. Vertical error bars correspond to the experimental uncertainty. The solid lines are linear fits yielding $n_{\text{c}}(0)=0.87\pm0.02\times10^{10}$~cm$^{-2}$. Top inset: Current-voltage characteristic measured at a temperature of 30~mK. Bottom inset: Arrhenius plots of the resistivity in the insulating phase for two electron densities. The densities in both insets are indicated in cm$^{-2}$.}
\label{fig1}
\end{figure}

The resistivity, $\rho$, as a function of temperature, $T$, is shown in Fig.~\ref{fig1}(a) for different electron densities, $n_{\text{s}}$, on both sides of the metal-insulator transition. While at the highest temperature the difference between the resistivities measured at the lowest and highest densities differ by less than two orders of magnitude, at the lowest temperature this difference exceeds six orders of magnitude. We identify the transition point at $n_{\text{c}}(0)=0.88\pm0.02\times10^{10}$~cm$^{-2}$, based on the ${\rm d}\rho/{\rm d}T$ sign-change criterion taking account of the tilted separatrix \cite{punnoose2005metal}. The low-temperature drop of the resistivity in the ultra-clean 2D electron system in SiGe/Si/SiGe quantum wells reaches a factor of about 12. This is the highest value reported in any 2D electron system \cite{melnikov2019quantum}.

The location of the MIT point can also be determined by studying the insulating side of the transition, where the resistance has an activated form, as shown in the bottom inset of Fig.~\ref{fig1}(b); note that the activation energy, $E_{\text a}$, can be determined provided $E_{\text a}>k_{\text{B}}T$. Figure~\ref{fig1}(b) shows the activation energy in temperature units, $E_{\text a}/k_{\text{B}}$, as a function of the electron density. Near the critical point, this dependence corresponds to the constant thermodynamic density of states and should be linear; the relative accuracy of determination of $E_{\text a}$ increases with increasing activation energy, and the linear fit should be drawn through all data points. The activation energy extrapolates to zero at $n_{\text{c}}(0)=0.87\pm0.02\times10^{10}$~cm$^{-2}$ which coincides, within the experimental uncertainty, with the value of $n_{\text{c}}$ determined from the temperature derivative criterion. Furthermore, in the insulating state, a typical low-temperature $I$-$V$ curve shows a step-like function: the voltage rises abruptly at low currents and then almost saturates, as seen in the top inset of Fig.~\ref{fig1}(b). The magnitude of the step is $2\, V_{\text{c}}$, where $V_{\text{c}}$ is the threshold voltage. The threshold behavior of the $I$-$V$ curves has been explained \cite{polyakov1993conductivity,shashkin1994insulating} within the concept of the breakdown of the insulating phase that occurs when the localized electrons at the Fermi level gain enough energy to reach the mobility edge in an electric field, $V_{\text{c}}/d$, over a distance of the localization length, $L$ (here $d$ is the distance between the potential probes). The values $E_{\text a}/k_{\text{B}}$ and $V_{\text{c}}$ are related via the localization length, which is temperature-independent and diverges near the transition as $L(E_{\text{F}})\propto (E_{\text{c}}-E_{\text{F}})^{-s}$ with exponent $s$ close to unity \cite{shashkin1994insulating} (here $E_{\text{c}}$ is the mobility edge and $E_{\text{F}}$ is the Fermi level). This corresponds to a linear dependence of the square root of $V_{\text{c}}$ on $n_{\text{s}}$ near the MIT, as seen in Fig.~\ref{fig1}(b). The dependence extrapolates to zero at the same electron density as $E_{\text a}/k_{\text{B}}$. A similar analysis has been previously performed \cite{shashkin2001metal} in a 2D electron system in Si MOSFETs and has yielded similar results, thus adding confidence that the MIT in 2D is a genuine quantum phase transition.

The critical electron density for the MIT is almost an order of magnitude smaller than that in the least-disordered Si MOSFETs, where $n_{\text{c}}(0)\approx 8\times10^{10}$~cm$^{-2}$. Such a difference can indeed be expected for an interaction-driven MIT. The interaction parameter, $r_{\text{s}}$, is defined as the ratio of the Coulomb and Fermi energies, $r_{\text{s}}=g_{\text{v}}/(\pi n_{\text{s}})^{1/2}a_{\text{B}}$, where $g_{\text{v}}=2$ is the valley degeneracy and $a_{\text{B}}$ is the effective Bohr radius in the semiconductor. We compare the value of the interaction parameter at the critical density $n_{\text{c}}$ in SiGe/Si/SiGe quantum wells with that in Si MOSFETs (where $r_{\text{s}}\approx20$). The two systems differ by the level of the disorder, the thickness of the 2D layer, and the dielectric constant (7.7 in Si MOSFETs and 12.6 in SiGe/Si/SiGe quantum wells). Due to the higher dielectric constant, the interaction parameter at the same electron density is smaller in SiGe/Si/SiGe quantum wells by approximately 1.6. In addition, the effective $r_{\text{s}}$ value is reduced further due to the much greater thickness of the 2D layer in the SiGe/Si/SiGe quantum wells, which results in a smaller form-factor \cite{ando1982electronic}. Assuming that the effective mass in the SiGe barrier is $\approx0.5\, m_{\text{e}}$ and estimating the barrier height at $\approx25$~meV, we evaluate the penetration of the wave function into the barrier and obtain the effective thickness of the 2D layer $\approx200$~\AA\ compared to $\approx50$~\AA\ in Si MOSFETs. This yields the additional suppression of $r_{\text{s}}$ in the SiGe/Si/SiGe quantum wells compared to Si MOSFETs by a factor of about 1.3. Thus, the electron densities $n_{\text{c}}$ correspond to $r_{\text{s}}\approx20$ in both Si MOSFETs and SiGe/Si/SiGe quantum wells, which is consistent with the results of Ref.~\cite{shashkin2007strongly}.

\section{Quantum phase transition in spinless electron system}

\begin{figure}
\scalebox{.95}{\includegraphics[width=\columnwidth]{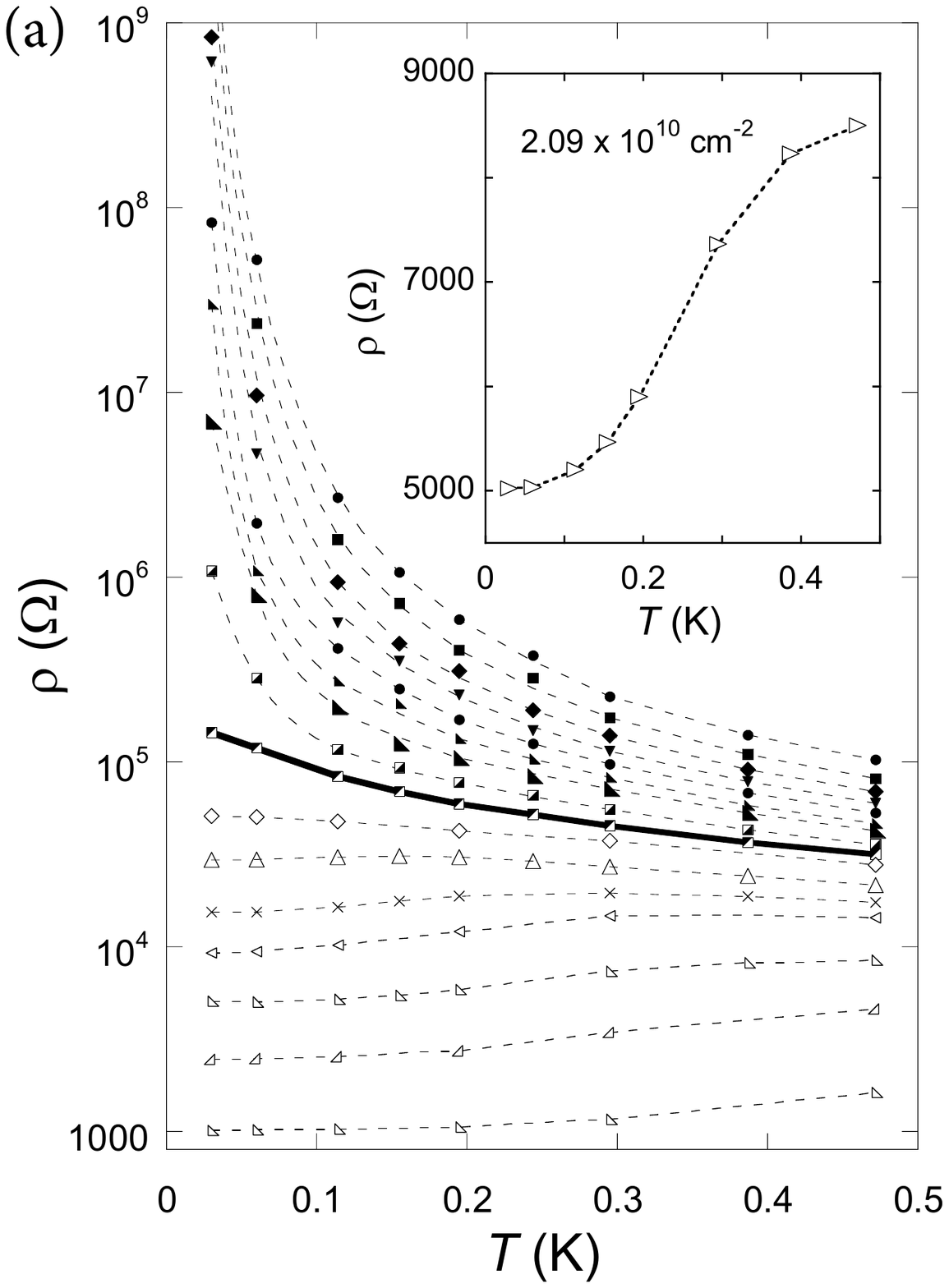}}
\scalebox{.95}{\includegraphics[width=\columnwidth]{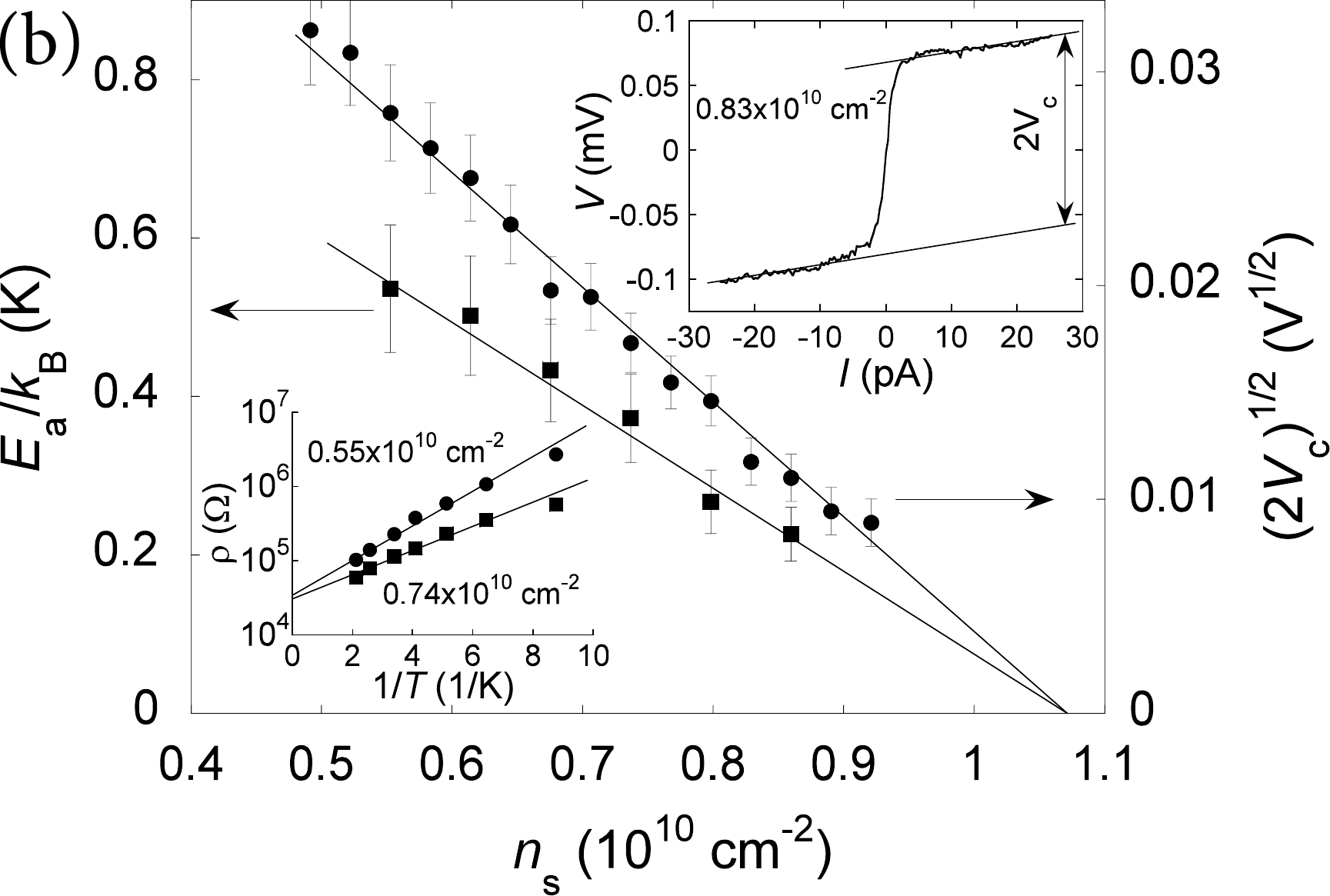}}
\caption{(a) Resistivity of an electron system in a SiGe/Si/SiGe quantum well placed in the spin-polarizing magnetic field $B^*$ as a function of temperature for electron densities (from top to bottom) 0.55, 0.61, 0.68, 0.74, 0.80, 0.86, 0.92, 1.01, 1.11, 1.22, 1.35, 1.54, 1.72, 2.09, 2.64, and $3.75\times 10^{10}$~cm$^{-2}$. The solid line corresponds to the separatrix. The magnetic fields used are spanned in the range between approximately 1 and 2~T.  The inset shows a closeup view of $\rho(T)$ for $n_{\text s}=2.09\times10^{10}$~cm$^{-2}$. (b) Activation energy, $E_{\text a}$, and square root of the threshold voltage, $V_{\text c}^{1/2}$, \textit{vs}.\ electron density.  Solid lines correspond to the best linear fits.  Upper inset: a typical $I$-$V$ dependence on the insulating side of the MIT at $T=30$~mK.  Lower inset: Arrhenius plots of the temperature dependence of the resistivity for two electron densities on the insulating side.}
\label{fig2}
\end{figure}

The electron spectrum in silicon-based 2D systems contains two almost degenerate valleys, which is expected to promote metallicity \cite{punnoose2001dilute,punnoose2005metal,fleury2008many}. Therefore, the metallic state may, in principle, survive in these systems in the presence of spin-polarizing magnetic fields.  Below, we show that the metallic state in ultra-low-disorder SiGe/Si/SiGe quantum wells survives even when the electrons' spins are fully polarized \cite{melnikov2020metallic}.

In Fig.~\ref{fig2}(a) we plot the resistivity $\rho(T)$, measured in a parallel magnetic field strong enough to polarize the electron spins, for different electron densities.  The magnetic field for complete spin polarization $B^*$ is density-dependent and was determined by the saturation of the $\rho(B_\parallel)$ dependence, which corresponds to the lifting of the spin degeneracy \cite{okamoto1999spin,vitkalov2000small}; the magnetic fields used in our experiments fell within the range between approximately 1 and 2~T. At the lowest temperatures, the resistivity exhibits a strong metallic temperature dependence ($d\rho/dT>0$) for electron densities above a certain critical value, $n_{\text c}(B^*)$, and an insulating behavior ($d\rho/dT<0$ with resistivity diverging as $T\rightarrow0$) for lower densities.  Assuming that the extrapolation of $\rho(T)$ to $T=0$ is valid and taking into account that the curve separating metallic and insulating regimes should be tilted \cite{punnoose2005metal}, we identify the critical density for the metal-insulator transition $n_{\text c}(B^*)=(1.11\pm0.05)\times10^{10}$~cm$^{-2}$ in a way similar to the case of $B=0$ \cite{melnikov2019quantum}. The $\rho(T)$ dependences on the metallic side of the transition at $n_{\text s}$ just above the critical density are non-monotonic: while at temperatures exceeding a density-dependent value $T_{\text {max}}$, the derivative $d\rho/dT$ is negative (``insulating-like''), it changes sign at temperatures below $T_{\text {max}}$. The measurements were restricted to 0.5~K that is the highest temperature at which the saturation of the $\rho(B_\parallel)$ dependence could still be achieved; the restriction is likely to reflect the degeneracy condition for the dilute electron system with low Fermi energy.

On the metallic side of the transition ($n_{\text s}>n_{\text c}(B^*)$), the maximum resistivity drop with decreasing temperature below 0.5~K reaches almost a factor of 2 (see the inset in Fig.~\ref{fig2}(a)), which is weaker compared to more than an order-of-magnitude drop in this system at $B=0$ \cite{melnikov2019quantum}.  Still, the metallic temperature behavior of spinless electrons in SiGe/Si/SiGe quantum wells remains strong and similar to that observed in $p$-type GaAs/AlGaAs heterostructures in zero magnetic field \cite{hanein1998the,gao2006spin}.

Similarly to the way it was done in the previous section, one can also deduce the critical density for the MIT from two additional criteria not requiring the extrapolation of the data to $T=0$: namely, vanishing of the activation energy and nonlinearity of the current-voltage characteristics on the insulating side of the transition ($n_{\text s}<n_{\text c}(B^*)$).  The temperature dependences of the resistivity have an activation character on the insulating side in the vicinity of the transition (see the lower inset in Fig.~\ref{fig2}(b)); the activation energy $E_{\text a}$ is plotted in the main panel of Fig.~\ref{fig2}(b) as a function of $n_{\text s}$.  The dependence is linear, which corresponds to the constant thermodynamic density of states near the critical point, and extrapolates to zero at $n_{\text c}(B^*)=(1.07\pm0.03)\times10^{10}$~cm$^{-2}$ which coincides, within the experimental uncertainty, with the value of $n_{\text c}(B^*)$ determined from the temperature derivative criterion.

A typical $I$-$V$ curve measured on the insulating side of the MIT ($n_{\text s}<n_{\text c}(B^*)$) is shown in the upper inset to Fig.~\ref{fig2}(b).  The $V(I)$ dependence obeys Ohm's law in a very narrow interval of currents $\left|I\right|\lesssim1$~pA and almost saturates at higher excitation currents. The dependence $V_{\text c}^{1/2}(n_{\text s})$ is linear near the MIT and extrapolates to zero at the same electron density as the $E_{\text a}(n_{\text s})$ dependence.

The fact that the spinless electrons behave differently as compared to those in Si MOSFETs, where the metallic temperature behavior of the resistance is completely quenched, can be attributed to different intervalley scattering rates \cite{melnikov2020metallic}. The level of short-range disorder potential in our samples is some two orders of magnitude lower than that in the least disordered Si MOSFETs, hence the intervalley scattering rate should be small compared to that in Si MOSFETs, corresponding to the case of two distinct valleys. The critical electron density $n_{\text c}(B^*)$ for the MIT in the spinless electron system is higher by a factor of about 1.2 compared to $n_{\text{c}}(0)$ in zero magnetic field, which is consistent with theoretical calculations \cite{dolgopolov2017spin}. The fact that the observed metallic temperature behavior is comparable to that in strongly interacting, spin-unpolarized single-valley 2D systems in the cleanest $p$-type GaAs/AlGaAs heterostructures indicates the same role of spins and distinct valleys with respect to the existence of the metallic state and the MIT.

\section{Spin effect on the low-temperature resistivity maximum}

Early theories of the metallic state in strongly interacting 2D systems \cite{finkelstein1983influence,finkelstein1984weak,castellani1984interaction} were focused on the interplay between disorder and interactions using renormalization-group scaling theory.  Later, the theory was extended to account for the existence of multiple valleys in the electron spectrum \cite{punnoose2001dilute,punnoose2005metal}.  At temperatures well below the Fermi temperature, the resistivity was predicted to grow with decreasing temperature, reach a maximum at $T=T_{\text{max}}$, and then decrease as $T\rightarrow0$.  The maximum in $\rho(T)$ dependence corresponds to the temperature at which the interaction effects become strong enough to stabilize the metallic state and overcome the quantum localization.  This theoretical prediction, which is applicable only within the so-called diffusive regime (roughly, $k_{\text B}T\tau/\hbar<1$, where $\tau$ is the mean-free time), was found to be consistent with the experimental $\rho(T)$ data in silicon MOSFETs \cite{punnoose2001dilute,punnoose2010test,anissimova2007flow}, but only in a narrow range of electron densities near $n_{\text c}(0)$ for the resistivities low compared to $\pi h/e^2$.  However, strong temperature dependence of the resistivity has been experimentally observed in a wide range of electron densities: up to five times the critical density, including the so-called ballistic regime (roughly, $k_{\text B}T\tau/\hbar>1$), where the renormalization-group scaling theory is not relevant.

An alternative interpretation of the temperature dependence of the resistivity is based on the so-called Wigner-Mott scenario, which focuses on the role of strong electron-electron interactions. The simplest theoretical approach to non-perturbatively tackle the interactions as the main driving force for the MIT is based on dynamical mean-field theory (DMFT) methods \cite{camjayi2008coulomb,radonjic2012wigner,dobrosavljevic2017wigner} using the Hubbard model at half-filling.  On the metallic side near the MIT, the resistivity is predicted to initially increase as the temperature is reduced, reach a maximum, $\rho_{\text {max}}$, at a temperature $T_{\text {max}}$, and then decrease as $T\rightarrow0$ so that the resistivity change $\rho(T)-\rho(0)$, normalized by its maximum value, is a universal function of $T/T_{\text {max}}$. According to this theory, $T_{\text {max}}$ corresponds to the quasiparticle coherence temperature, which is of the order of the Fermi temperature $T_{\text F}$ determined by the effective electron mass renormalized by interactions: $T_{\text {max}}\sim T_{\text F}$. Below this temperature, the elastic electron-electron scattering corresponds to coherent transport, while at higher temperatures, the inelastic electron-electron scattering becomes strong and gives rise to a fully incoherent transport. Notably, similar functional form of the resistivity $\rho(T)$ can be expected within the screening theory in its general form (for more on this, see Ref.~\cite{shashkin2020manifestation}). Similar non-monotonic $\rho(T)$ dependence with a maximum at $T_{\text {max}}\sim T_{\text F}$ is predicted by another approach based on the Pomeranchuk effect expected within a phase coexistence region between the Wigner crystal and a Fermi liquid \cite{spivak2003phase,spivak2004phases,spivak2006transport}. The first two approaches allow for a quantitative comparison with the experiment.

\begin{figure}
\scalebox{.95}{\includegraphics[width=\columnwidth]{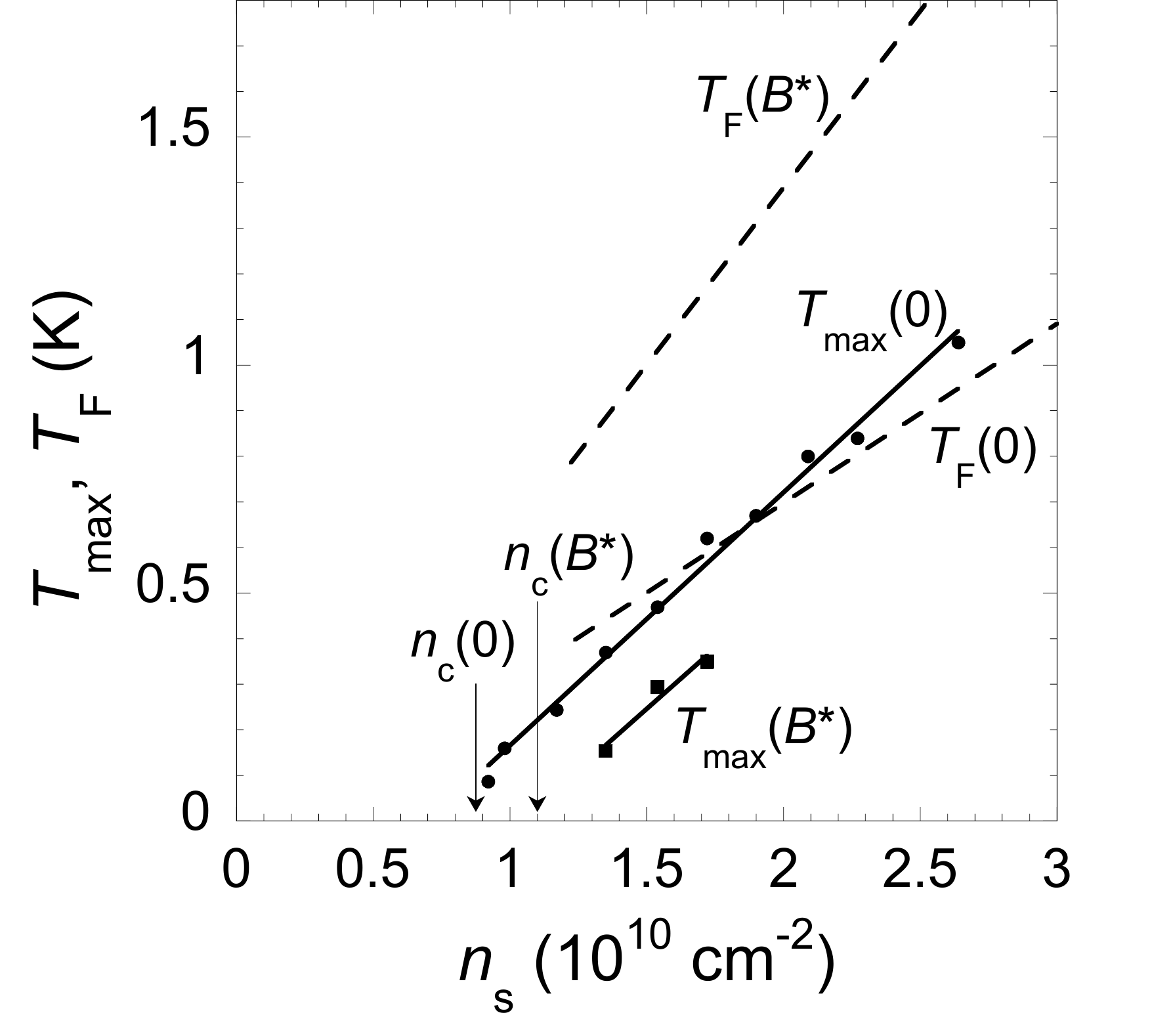}}
\caption{$T_{\text{max}}$ as a function of electron density in $B=0$ (circles) and in $B=B^*$ (squares). Solid lines are linear fits. Critical electron densities for the metal-insulator transition in $B=0$ and $B=B^*$ are indicated. Dashed lines show the Fermi temperatures $T_{\text F}$ in $B=0$ and $B=B^*$ calculated using the low-temperature value $B^*$ and Eq.~(\ref{eq1}), see text.}
\label{fig3}
\end{figure}

In Fig.~\ref{fig3} we plot the values of $T_{\text{max}}$ as a function of the electron density in $B=0$ and $B=B^*$. The data for $T_{\text{max}}(B^*)$ lie significantly lower than those for $T_{\text{max}}(0)$. Interestingly, each dependence can be described by a linear function that extrapolates to zero at $n_{\text s}$ close to $n_{\text c}(0)$ or $n_{\text c}(B^*)$, and the slopes of both dependences are close to each other. We also plot the calculated values of renormalized Fermi temperatures $T_{\text F}$ for both cases. In zero magnetic field, the density dependences of the resistivity maximum temperature $T_{\text{max}}(0)$ and the Fermi temperature $T_{\text F}(0)$ are close to each other in the electron density range where they overlap. However, there is a qualitative difference between the behavior of $T_{\text{max}}$ and that of $T_{\text F}$ when lifting the spin degeneracy $g_{\text s}$. Rather than increasing along with the Fermi temperature, the value $T_{\text{max}}$ decreases when polarizing electron spins \cite{shashkin2022spin}.

The Fermi temperature $T_{\text F}(B^*)$ has been calculated from the low-temperature value $B^*$ based on the equality of the Fermi energy of completely spin-polarized electrons and the Zeeman energy in the polarization field $B^*$ \cite{melnikov2017indication}:
\begin{equation}
k_{\text B}T_{\text F}(B^*)=\frac{h^2n_{\text s}}{2\pi g_{\text v}m}=g_{\text F}\mu_{\text B}B^*,\label{eq1}
\end{equation}
where $k_{\text B}$ is the Boltzmann constant, $g_{\text v}=2$ is the valley degeneracy, $m$ is the renormalized energy-averaged effective mass that is determined by the density of states, $g_{\text F}\approx g_0=2$ is the $g$-factor at the Fermi level, $g_0$ is the $g$-factor in bulk silicon, and $\mu_{\text B}$ is the Bohr magneton. We argue that the Fermi temperature $T_{\text F}(0)$ of spin-unpolarized electrons is approximately half of the Fermi temperature $T_{\text F}(B^*)$ of completely spin-polarized ones. Indeed, it was experimentally shown in Ref.~\cite{shashkin2006pauli} that the electron spin magnetization is proportional to the parallel magnetic field in the range up to $B=B^*$ for the clean, strongly interacting 2D electron system in Si MOSFETs that is similar to the 2D electron system in SiGe/Si/SiGe quantum wells.  (For strongly disordered Si MOSFETs, the band tail of localized electrons persists into the metallic regime \cite{prus2003thermodynamic} in which case both the nonlinear magnetization as a function of parallel magnetic field and the shift of the dependence $B^*(n_{\text s})$ to higher densities are observed due to the presence of localized electron moments in the band tail \cite{dolgopolov2002comment,gold2002on,shashkin2005metal,teneh2012spin,pudalov2018probing}.) Taking into account the smallness of the exchange effects in the 2D electron system in silicon so that the $g$-factor is approximately constant close to $g_0=2$ at low densities \cite{kravchenko2004metal,shashkin2005metal,melnikov2017indication}, this indicates that the renormalized density of states in a spin subband is approximately constant below the Fermi level, independent of the magnetic field. Therefore, the change of $T_{\text F}$ when lifting the spin degeneracy should be controlled by the change of $g_{\text s}$. As concerns the band flattening corresponding to a peak in the density of states at the Fermi level, observed in the 2D electron system in SiGe/Si/SiGe quantum wells, the Fermi energy is not particularly sensitive to this flattening, at least, not too close to the critical point \cite{melnikov2017indication}. So, one expects that the relation $T_{\text F}(0)\approx T_{\text F}(B^*)/2$ holds for the data in question. We stress that its accuracy is not crucial for our qualitative results.

The dynamical mean-field theory successfully describes the closeness of $T_{\text{max}}$ and the renormalized Fermi temperature $T_{\text F}$ in zero magnetic field, as well as the resistivity drop at temperatures below $T_{\text{max}}$ in both spin-unpolarized and fully spin-polarized electron systems (see the next section). However, the observed decrease of $T_{\text{max}}$ when lifting the spin degeneracy is opposite to the predictions of DMFT. At the same time, the reduced value of $T_{\text{max}}$ in spin-polarizing magnetic fields is consistent with the predictions of the renormalization-group scaling theory, but $T_{\text{max}}$ in zero magnetic field is in disagreement with this theory. The observed behavior of $T_{\text{max}}$ cannot be described by existing theories. Nor can it be explained in terms of the increase of the residual disorder potential and the reduction of the electron interaction strength due to the reduced spin degrees of freedom in spin-polarizing magnetic fields, because the relation $T_{\text {max}}\sim T_{\text F}$ still holds for clean Si MOSFETs \cite{radonjic2012wigner,dobrosavljevic2017wigner} and low-mobility Si/SiGe quantum wells \cite{lu2011termination} in zero magnetic field \cite{shashkin2022spin}. This indicates the spin-related origin of the effect.

\section{Scaling of the non-monotonic $\rho(T)$}

\begin{figure}[b]
\scalebox{.95}{\includegraphics[width=\columnwidth]{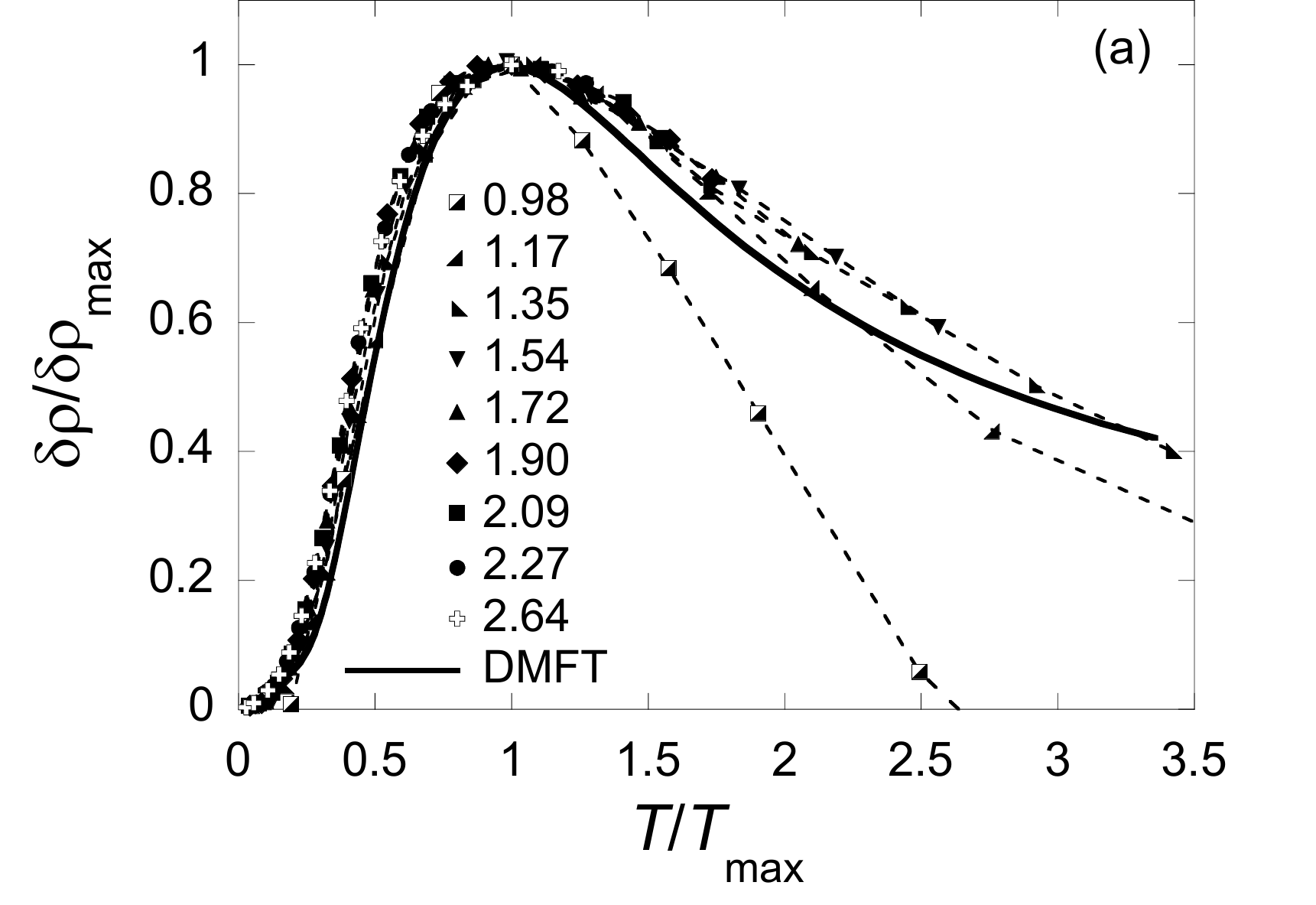}}
\scalebox{.95}{\includegraphics[width=\columnwidth]{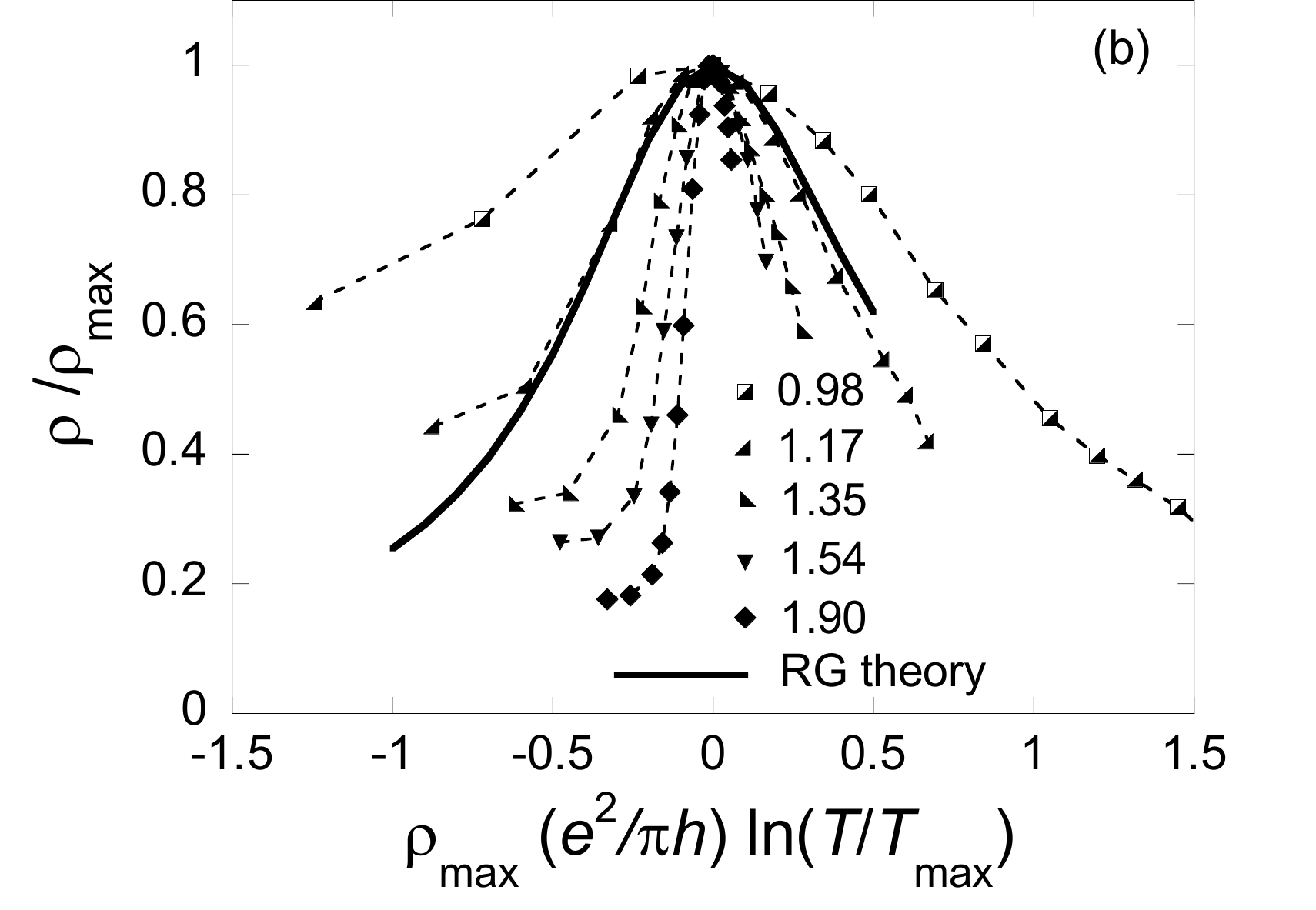}}
\caption{(a) The ratio $(\rho(T)-\rho(0))/(\rho_{\text {max}}-\rho(0))$ as a function of $T/T_{\text {max}}$ in $B=0$.  The solid line shows the result of the dynamical mean-field theory in the weak-disorder limit \cite{camjayi2008coulomb,radonjic2012wigner,dobrosavljevic2017wigner}. The electron densities are indicated in units of $10^{10}$~cm$^{-2}$. (b) The ratio $\rho/\rho_{\text {max}}$ as a function of the product $\rho_{\text {max}} \ln(T/T_{\text {max}})$.  The solid line is the result of the renormalization-group scaling theory \cite{punnoose2001dilute,punnoose2005metal}.
}
\label{fig4}
\end{figure}

The results of the scaling analysis of the data in $B=0$ in the spirit of the dynamical mean-field theory \cite{camjayi2008coulomb,radonjic2012wigner,dobrosavljevic2017wigner} are shown in Fig.~\ref{fig4}(a).  The data scale perfectly in a wide range of electron densities and are described well by the theory in the weak-disorder limit \cite{shashkin2020manifestation}; we emphasize that at some electron densities, the changes of the resistivity with temperature exceed an order of magnitude.  Deviations from the theoretical curve arise in the high-temperature limit in the transient region and become pronounced for $T>T_{\text {max}}$ at electron densities within $\sim10$\% of the critical value, which in these samples is close to $n_{\text c}(0)\approx 0.88\times10^{10}$~cm$^{-2}$.

\begin{figure}[b]
\scalebox{.92}{\includegraphics[width=\columnwidth]{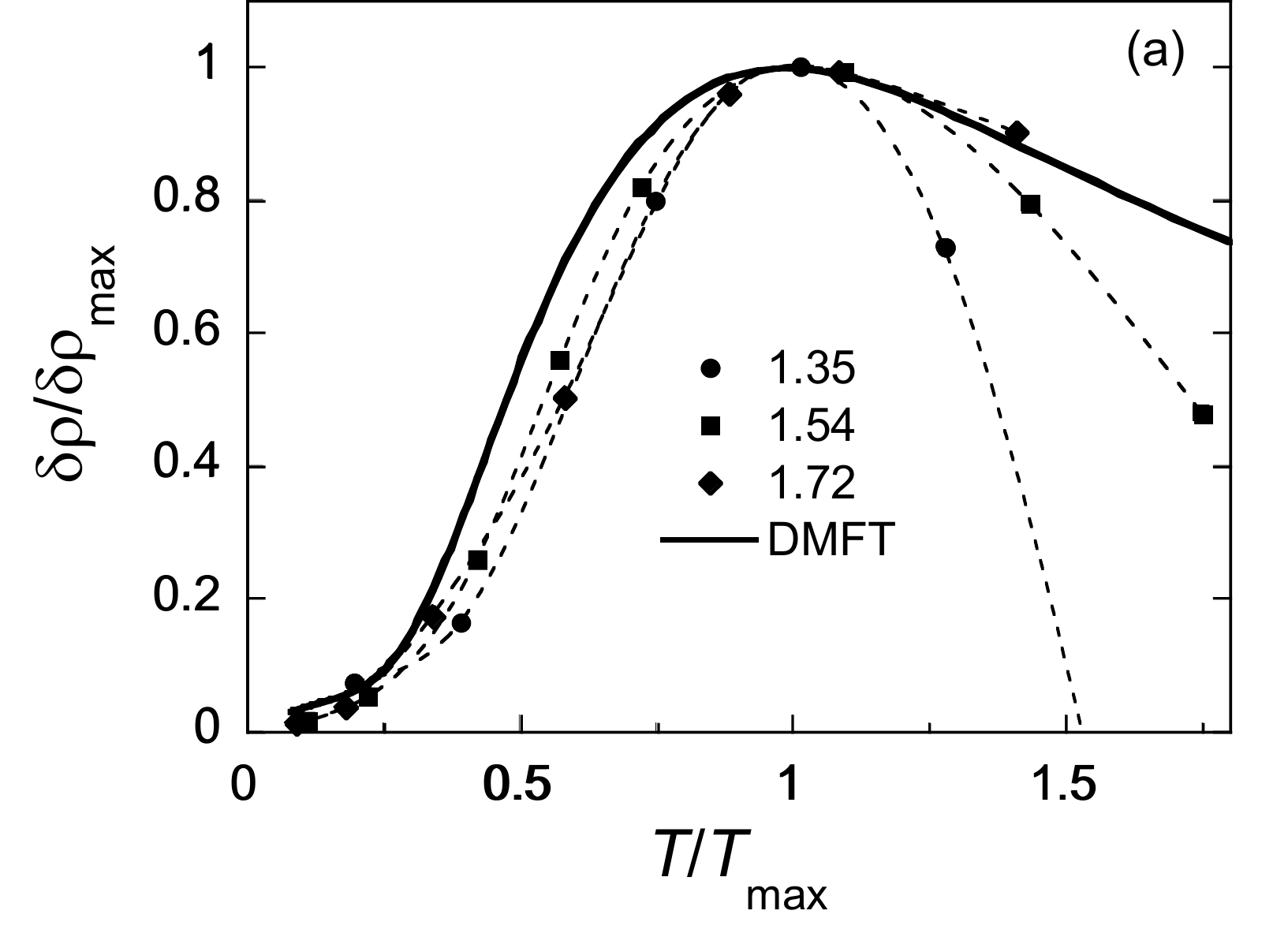}}
\scalebox{.92}{\includegraphics[width=\columnwidth]{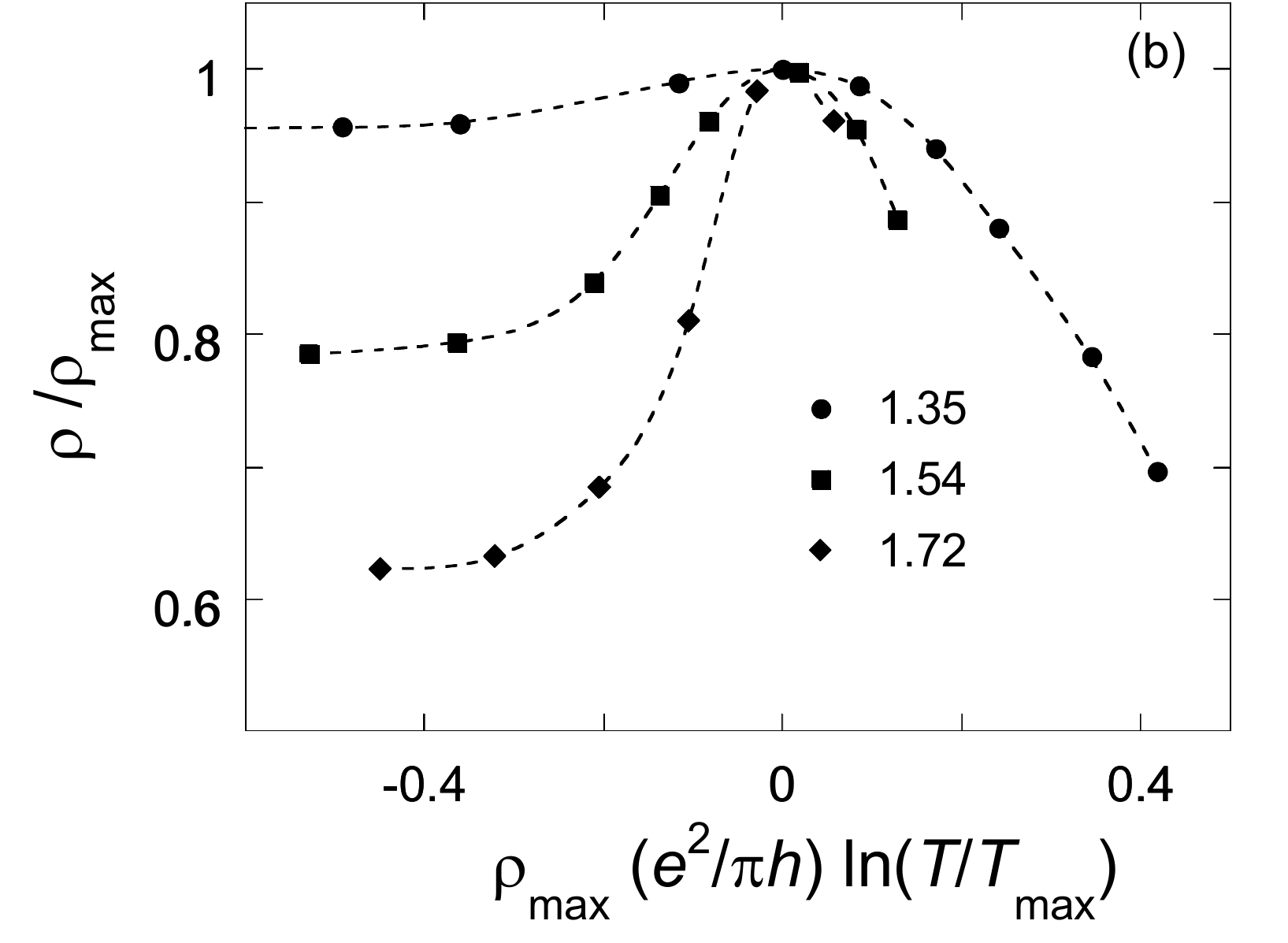}}
\caption{(a) The ratio $\delta\rho/\delta\rho_{\text{max}}$ plotted as a function of $T/T_{\text{max}}$ in $B=B^*$. The solid line is the result of DMFT in the weak-disorder limit \cite{camjayi2008coulomb,radonjic2012wigner,dobrosavljevic2017wigner}. The electron densities are indicated in units of $10^{10}$~cm$^{-2}$. (b) The analysis based on the scaling form suggested by the renormalization-group scaling theory \cite{punnoose2001dilute,punnoose2005metal}.}
\label{fig5}
\end{figure}

For proper perspective and comparison, we also show the results of a scaling analysis done in accordance with the renormalization-group scaling theory \cite{punnoose2001dilute,punnoose2005metal}, according to which the normalized resistivity $\rho/\rho_{\text {max}}$ should be a universal function of the product $\rho_{\text {max}}\ln(T/T_{\text {max}})$.  The results are plotted in Fig.~\ref{fig4}(b).  Only the data obtained at $n_{\text s}=1.17\times10^{10}$~cm$^{-2}$ coincide nearly perfectly with the theoretical curve, although some deviations occur at the lowest temperature.  Pronounced deviations from the theory are evident at both higher and lower $n_{\text s}$.  At lower electron densities, the scaled experimental curves become wider than the theoretical one, and at higher densities, they become narrower.  A similar shrinkage of the scaled curves with increasing $n_{\text s}$ was reported earlier in Refs.~\cite{punnoose2001dilute,anissimova2007flow,radonjic2012wigner}.  One should take into account, however, that theory \cite{punnoose2001dilute,punnoose2005metal} has been developed for 2D electron systems that, on the one hand, are in the diffusive regime and, on the other hand, their resistivities are low compared to $\pi h/e^2$: at higher values of $\rho$, higher-order corrections become important and cause deviations from the universal scaling curve.  As a result, the applicable range of parameters becomes very narrow.

In Fig.~\ref{fig5}(a), we plot the ratio $\delta\rho/\delta\rho_{\text{max}}=(\rho(T)-\rho(0))/(\rho(T_{\text{max}})-\rho(0))$ as a function of $T/T_{\text{max}}$ in $B=B^*$ so as to check the applicability of the DMFT for spinless electrons. The curve for the highest electron density follows the theoretical dependence in the weak-disorder limit at all temperatures. Two other curves for lower electron densities also follow the theoretical dependence at $T\leq T_{\text{max}}$ but deviate from the theory at higher temperatures, revealing the behavior similar to that observed at low $n_{\text s}$ in zero magnetic field \cite{shashkin2020manifestation}. Albeit the density range of the applicability of DMFT to the completely spin-polarized system is not as wide as that in $B=0$, the low-temperature resistivity drop is described by the theory, similar to the case of the spin-unpolarized electron system. In Fig.~\ref{fig5}(b) we plot the ratio $\rho/\rho_{\text{max}}$ in the fully spin-polarized system as a function of $\rho_{\text{max}}(e^2/\pi h)\,{\text{ln}}(T/T_{\text{max}})$, which is the scaling form suggested by the renormalization-group scaling theory \cite{punnoose2001dilute,punnoose2005metal}. The data do not scale in the range of electron densities studied.  Thus, the $\rho(T)$ data are best described by DMFT.

\section{Conclusions}

We have found that the metal-insulator transition occurs in the strongly-interacting two-valley two-dimensional electron system in ultra-high mobility SiGe/Si/SiGe quantum wells in zero magnetic field and survives in the spinless system in spin-polarizing magnetic fields. In both cases, this is accompanied by the non-monotonic temperature-dependent resistivity on the metallic side near the transition. In zero magnetic field, the resistivity maximum temperature is found to be close to the renormalized Fermi temperature. However, rather than increasing along with the Fermi temperature, the value $T_{\text{max}}$ decreases appreciably for spinless electrons in spin-polarizing magnetic fields. The observed behavior of $T_{\text{max}}$ cannot be described by existing theories. The results indicate the spin-related origin of the effect. At the same time, the low-temperature resistivity drop in both spin-unpolarized and spinless electron systems is described quantitatively by the dynamical mean-field theory.

\section{Acknowledgments}

A.A.S. was supported by RSF Grant No.\ 22-22-00333. S.V.K. was supported by NSF Grant No.\ 1904024.



\end{document}